\documentclass{journal}

\usepackage{graphicx} 
\usepackage[utf8]{inputenc}
\usepackage[T1]{fontenc}
\usepackage{amsmath}      
\usepackage{cleveref}     
\usepackage{url}          
\usepackage{footnote}     
\usepackage{float}
\usepackage{authblk}
\usepackage{geometry}

\title{\Large Vast TVB parameter space exploration: \\ A Modular Framework for Accelerating the Multi-Scale Simulation of Human Brain Dynamics} 

\author[1]{Michiel van der Vlag $^*$}
\author[2]{Lionel Kusch}
\author[3]{Alain Destexhe}
\author[2]{Viktor Jirsa}
\author[1]{Sandra Diaz-Pier}
\author[3]{Jennifer Goldman $^*$}

\affil[1]{\small SDL Neuroscience, Institute for Advanced Simulation, Jülich Supercomputing Centre (JSC), Jülich Research Center and JARA, Germany}
\affil[2]{\small Institut de Neurosciences des Syst\`emes, Aix-Marseille University, INSERM, Marseille, France}
\affil[3]{\small Paris Saclay University, Institute of Neuroscience, CNRS, Gif-sur-Yvette, France}
\affil[*]{Correspondence: m.van.der.vlag@fz-juelich.de, jennifer.goldman.mcgill@gmail.com}

\date{November 2023}

\begin{document}

\maketitle

\abstract{

Global neural dynamics emerge from multi-scale brain structures, with neurons communicating through synapses to form transiently communicating networks. 
Network activity arises from intercellular communication that depends on the structure of connectome tracts and local connection, intracellular signalling cascades, and the extracellular molecular milieu that regulate cellular properties. 
Multi-scale models of brain function have begun to directly link the emergence of global brain dynamics in conscious and unconscious brain states to microscopic changes at the level of cells. 
In particular, AdEx mean-field models representing statistical properties of local populations of neurons have been connected following human tractography data to represent multi-scale neural phenomena in simulations using The Virtual Brain (TVB). 
While mean-field models can be run on personal computers for short simulations, or in parallel on high-performance computing (HPC) architectures for longer simulations and parameter scans, the computational burden remains high and vast areas of the parameter space remain unexplored. 
In this work, we report that our TVB-HPC framework, a modular set of methods used here to implement the TVB-AdEx model for GPU and analyze emergent dynamics, notably accelerates simulations and substantially reduces computational resource requirements. 
The framework preserves the stability and robustness of the TVB-AdEx model, thus facilitating finer resolution exploration of vast parameter spaces as well as longer simulations previously near impossible to perform.  
Comparing our TVB-HPC framework to the TVB-AdEx, we first show similarity of parameter combinations giving rise to patterns of functional connectivity between brain regions.
Varying global coupling together with spike-frequency adaptation, we next replicate their inter-dependence in inducing transitions between dynamics associated with conscious and unconscious brain states. 
Further exploring parameter space, we report a nonlinear interplay between the spike-frequency adaptation and subthreshold adaptation, as well as previously unappreciated interactions between the global coupling, adaptation, and propagation velocity of action potentials along the human connectome. 
Given that simulation and analysis toolkits are made public as open-source packages, our framework serves as a template onto which other models can be easily scripted and personalized datasets can be used for studies of inter-individual variability of parameters related to functional brain dynamics. 
These results thus represent potentially impactful, publicly-available methods for simulating and analyzing human brain states.}

\keywords{brain simulation; multi-scale; consciousness; high performance computing, mean-field model; connectome; brain dynamics; phase transition; brain state}

\newgeometry{margin=3.cm}

\section{Introduction}

The brain is a multi-scale organ, with relevant scales spanning molecules to synapses, to neurons, to networks or populations of neurons, to global brain anatomy, connectivity, and dynamics. 
Computational models that connect these biologically relevant scales are an active area of research \cite{goldman2019bridging}. 
However, a scale-integrated understanding - relating the actions and nonlinear interactions of microscopic variables to macroscopic observables underlying global brain function - remains empirically challenging to derive from multi-modal experimental measurements \cite{stefanovski2019linking}. 
Thus, computational models can support progress to bridge our understanding across spatio-temporal scales, testing models by replicating empirical data and eventually making predictions for future experimental studies. 
Recent progress has been made modeling global phase transitions between unconscious-like (synchronous, regular) and conscious-like (asynchronous, irregular) dynamics that emerge from the same human connectomes, depending on the parameters describing cellular phenomena at the level of spikes \cite{Goldman2023, Aquilu2023, neurolibCakan2021}. Specifically, human, mouse, and primate TVB-AdEx models can describe the emergence of conscious-like asynchronous, irregular activity at global brain scales, based on diminished spike-frequency adaptation in simulated waking states due to enhanced acetylcholine concentrations (common to conscious brain states). 
In contrast, unconscious-like synchronous, regular global brain activity can be simulated by enhancing spike-frequency adaptation, biologically related to diminished neuromodulation during sleep \cite{Goldman2023, Aquilu2023}. 
Further, cellular hyperpolarization, a mechanism onto which multiple anesthetics converge, can lead to unconscious-like dynamics \cite{goldman2023asynchronous}. 
However, biophysically-based multi-scale models are complex, requiring substantial computational resources to represent the large parameter spaces that describe relevant observables. 
In addition to studying the behavior of simulations resulting from any one variable, it will be important to understand the ensemble of parameters and observables, as multiple parameters may act redundantly and/or interact non-linearly. 
While parameters can be estimated through inference \cite{schirner2018inferring, hashemi2020bayesian}, such methods could be complemented by a better understanding of the parameter spaces linked to these complex, multi-scale, biophysical human brain models, in order to assess whether previously unconsidered predictions naturally emerge. 
We have selected the TVB-AdEx model for this benchmarking study due to its significant computational requirements and its substantial scientific contribution to theoretical studies of consciousness. 
Nevertheless, it is important to note that the presented framework is not restricted to the AdEx model; other neural mass models can be employed in accelerated/ameliorated form, thanks to our modular design.

The present work describes the GPU implementation of the TVB-AdEx, a multi-scale model that describes the population statistics of excitatory and inhibitory AdEx neurons using mean-field approximations that can roughly represent human brain regions. 
The AdEx mean-field model was initially developed to describe the spontaneous and evoked dynamics of local populations of neurons with transfer-functions semi-analytically derived \cite{zerlaut2018modeling, divolo}. 
To form the TVB-AdEx, mesoscopic AdEx mean-fields are placed on and connected by empirical anatomical and diffusion magnetic resonance imaging (MRI) data. 
The CPU-based\footnote{\url{https://github.com/davidaquilue/TVBAdEx_ParSweep/tree/main}} version of the TVB-AdEx has previously been used to study scale-integrated mechanisms underlying the emergence of conscious- and unconscious-like brain state dynamics using a local PC, EBRAINS \footnote{\url{https://www.ebrains.eu/}}, an online aggregate of models \cite{Goldman2023}, and an MPI multi-node CPU setup to study limited compound parametrizations for five parameters known to be fundamental to the behaviour of this AdEx model \cite{Aquilu2023}. 

Here, we report that our GPU implementation of the TVB-AdEx model substantially ameliorates the original CPU instantiation, by accelerating model performance and facilitating post data-analysis. 
The TVB-HPC framework is not another platform; it represents an expanded framework built upon RateML, TVB's model generator\cite{rateml2022}.
RateML is performant, modular, reusable, and outperforms solutions such as neurolib\cite{neurolibCakan2021}, FastTVB\cite{fasttvb}, and Pyrates\cite{PyRates} in terms of magnitude of explorable parameter space and concurrent TVB instances. 
Specifically, our results show that it is an excellent compute accelerator for TVB-AdEx simulations allowing for high-dimensional explorations of the parameter space and the identification of parameters acting and interacting to bias brain state dynamics. 
This work highlights the importance of performant processing for complex brain models, such as biophysical multi-scale TVB-AdEx models, since the dimensionality of parameter space grows easily, quickly turning the problem intractable on previously existing simulation platforms.
By applying a wide range of accessible analysis methods, users are further enabled to explore observables describing simulated brain dynamics, including power spectral densities, exhibitory and inhibitory firing rates, quantification of up- and down- states, and structure-function relationships, using a repository where these tools are publicly available in EBRAINS\footnote{\url{https://wiki.ebrains.eu/bin/view/Collabs/rateml-tvb/Lab}}.
The TVB-HPC project, including the driver and GPU Adex model, analysis toolkit and graph script are also available at \footnote{\url{https://github.com/DeLaVlag/vast-TVB-space}}

We report that a single GPU, an NVIDIA A100 Tensor Core GPU with $40$ GB and CUDA version $12.2.0$, can be employed to explore roughly $17,000$ parameter combinations for any number of parameters combinations, simulating $68$ brain regions for $50,000$ simulations steps of $0.1$ms each. 
Theoretically, employing all $936$ of the GPUs within the Juwels Booster super-computing cluster\footnote{\url{https://apps.fz-juelich.de/jsc/hps/juwels/booster-overview.html}}, each with $4$ GPUs connected via NVLink3, would allow a simulation to simultaneously execute approximately $40$ million TVB instances, each configured with a distinct set of parameters. 
The execution of a simulation representing $5$ seconds of biological time requires roughly $8$ minutes to complete for this entire ensemble of instances.
Thus, this implementation represents a powerful, personalizable simulation framework to complement and extend multi-scale empirical brain research, that is built into EBRAINS tools intended to be useful for clinical researchers, neuroscientists and providers. 

\section{Materials and Methods}

The TVB-AdEx model uses mean-field modeling to integrate properties of excitatory and inhibitory AdEx neurons across scales. 
The populations are connected via a human connectome. 
It was observed that the macroscopic dynamics resembling brain activity emerge during simulation, with global conscious- and unconscious-like activity emerging from the tuning of parameters representing microscale phenomena \cite{Goldman2023, goldman2023asynchronous, Aquilu2023}. 
\subsection{Mean Field model}
The TVB-AdEx model is based on a mean field capturing the second order of the firing rate of an excitatory and inhibitory of adaptative exponential integrate and fire neurons \cite{brette_adaptive_2005} using a Master Equation formalism \cite{boustani_master_2009,zerlaut2018modeling} and the first order of the adaptative current of the excitatory population \cite{divolo}. 
It results in a system of differential equations describing the time evolution of the mean firing rate ($\nu$) of the two population ($\mu={e,i}$) and the variance ($c_{\lambda,\eta}$ where $\lambda =\eta$) and covariance ($c_{\lambda,\eta}$ where $\lambda\neq\eta$) of excitatory and inhibitory firing rate ($\lambda={e,i}$ and $\eta={e,i}$) and the mean adaptative current ($W_{\mu}$) for the excitatory population.
\begin{eqnarray}\label{mean-field_eq}
T\frac{d\nu_{\mu}}{dt}&=&(F_{\mu}-\nu_{\mu})+\frac{1}{2}c_{\lambda\eta}\frac{\partial^2 F_{\mu}}{\partial\nu_{\lambda}\partial\nu_{\eta}} \\
T\frac{dc_{\lambda\eta}}{dt}  &=& \delta_{\lambda\eta}\frac{F_{\lambda}(1/T-F_{\eta})}{N_{\lambda}}+(F_{\lambda}-\nu_{\lambda})(F_{\eta}-\nu_{\eta})
\nonumber \\
&&\quad \qquad \qquad \qquad  +\frac{\partial F_{\lambda}}{\partial \nu_{\mu}}c_{\eta\mu}+\frac{\partial F_{\eta}}{\partial \nu_{\mu}}c_{\lambda\mu}-2c_{\lambda \eta} \\
\frac{\partial W}{\partial t} & = & -W/\tau_w + b \nu_e + a_e \left(\mu_V(\nu_e, \nu_i, W) - E_{L,e}\right) \label{popadapt} 
\end{eqnarray}

Where $T$ is the time constant of the firing rate equations and covariance equations and $F_{\mu}$ is the transfer function. $\tau_{w}$, $b$ and $a$ are respectively the time constant, the spike-frequency and subthreshold of the adaptation, $E_{L}$ is the leak reversal potential and $\mu_V$ is an estimation of the average voltage of the excitatory population.

A stochastic equation is added to this set of differential equation for describing the random fluctuations of the mean firing rate, the noise. This equation of the noise is based on an Ornstein-Uhlenbeck process. The computation of this independent stochastic equation is done before the simulation and storied in memory to be accessible across time-steps.
\begin{equation}
    \tau_{OU}\frac{dOU_t}{dt} = (\mu - OU_t) + \sigma \, dW_t
\end{equation}
where $\mu$ (=0.0) is the mean of the noise, $\sigma$ (=1.0) is the variance of the noise and $W_t$ is a Wiener process. $\tau_{OU}$ represent the time constant of the noise. 

\subsection{Transfer Function}

The transfer function (TF) gives the instantaneous mean firing rate following the state of the each populations and the external inputs \cite{boustani_master_2009, zerlaut2018modeling,divolo}. 
\begin{equation}
    F_{\mu=\{e, i\}}=F_{\mu=\{e, i\}}({\nu_e}^{k,tot}+wNoise.\,OU_{t}^{k}+c_{F_{\mu,e}}, \nu_i^{k}+c_{F_{\mu,i}}, W_{\mu}^{k}) \text{ for brain region $k$.}
\end{equation}
Where ${\nu_e}^{k,tot} = g.\left(\displaystyle\sum_{68}^{j=1} u_{kj} \nu_{e_j}(t-\tau_{kj})\right)$ is the weighted sum of all firing rates of all brain regions, input to the region $k$. 
The weights ($u_{kj}$) are defined by the connectome. 
The delays ($\tau_{k,j}$) equals the track lengths between the region $k$ and region $j$ divided by the speed. 
$OU_{t}^{k}$ is the noise associated with the brain region (see above). $c_{F_{\mu,\lambda}}$ are constant excitatory or inhibitory input to the population.

One of the main originality of this formalism is that the mean-field depends on the specification of the transfer function $F_{\mu=\{e, i\}}$, which can be obtained according to a semi-analytical approach \cite{zerlaut2018modeling}, and thus potentially applicable to many different models \cite{Carlu2020}.  
In this approach, the transfer function is numerically fit to single-cell responses using an analytic template depending on three parameters:
the mean voltage ($\mu_V$), its standard deviation ($\sigma{V}$) and its time correlation decay time ($\tau_{V}$). 
Additionally a voltage threshold ($V^{eff}_{thre}$) is estimated from the properties of an individual neuron taken as a second order polynomial, depicted in \cref{tab:polyc}), of the previous quantities ($\mu_V,\sigma{V},\tau_{V}$). 
Using the assumption that the voltage membrane of the population follows a normal law and the phenomenological voltage threshold, it is possible to estimate the mean firing rate of the population, i.e. the output of the transfer function\cite{divolo, zerlaut2018modeling}. 

\begin{table}[]
\begin{tabular}{|l|l|l|l|l|l|l|l|l|l|l|}
\hline
Cell Type & $P_0$ & $P_{\mu V}$ & $P_{\sigma V}$ & $P_{\tau N V}$ & $P_{\mu 2}$ & $P_{\sigma 2}$ & $P_{\tau N V 2}$ & $P_{V \sigma}$ & $P_{\mu V_\tau N}$ & $P_{\sigma V \tau N}$ \\ \hline
RS-Cell   & -48.9 & 5.1      & -25.0  & 1.4   & -0.41   & 10.5  & -36.0        & 7.4   & 1.2     & 40.7    \\ \hline
FS-Cell   & -51.4 & 0.4       & -8.3  & 0.2   & -0.5   & 1.4  & -14.6        & 4.5    & 2.8     & 15.3    \\ \hline
\end{tabular}
\caption{Polynomial of the phenomenological voltage threshold for the transfer function (mV)}
\label{tab:polyc}
\end{table}

To advance beyond previously explored corners of the phase space for the multi-scale TVB-AdEx model with the published implementation \cite{Aquilu2023, Goldman2023}, prohibitively large computational resources would have been required. 
Therefore, we have implemented the TVB-AdEx model for GPU, which can be more efficient for computations that are embarrassingly parallel.
Parameterization in general is embarrassingly parallel due an absence of dependency between simulations; each parametric combination does not depend on other concurrent simulations. 
GPU is therefore an excellent choice for acceleration; the data latency of the individual simulations is largely covered by the many concurrent simulations. 
The many smaller computational units of the GPU, in comparison to a CPU, have also proven to be sufficient for TVB simulations.

\subsection{TVB-AdEx to GPU}
To accommodate the intricate regimes associated with various parameter settings within the TVB-AdEx model, a GPU-accelerated version has been developed.
The TVB model generator RateML, offers the capability to create customized rate-based or mean-field models.
In the standard workflow of this tool, an XML file can be populated with TVB generic properties including the derivative functions of the model.
This file is then translated into a fully-fledged TVB model or GPU model and driver for the execution of the model. 
Because the AdEx model has a complex transfer function, describing the firing rate of the neuron populations, the regular usage of RateML is not possible. 
However, the GPU model introduced here, is based on the blueprint for TVB GPU models, manually annotated with the added model complexity including the function calls to the transfer function ($zerlaut.c$). 
The transfer function itself is specified in a separate header file ($zerlaut.h$). 
Next to a GPU model which supports parameter sweeps, also the model driver ($model\_driver\_zerlaut.py$) object is generated when invoking RateML. 
Furthermore, this driver has been manually annotated for the computation and comparison of the functional connectivity (FC) to an externally obtained FC.

The model driver and model for the an TVB simulation using the GPU AdEx can be initiated on any CUDA capable GPU device. An example of the execution command follows: 

\begin{verbatim}
    python model_driver_zerlaut_mpi.py -s0 6 -s1 6 -s2 6 -s3 6 -s4 6 
    -n 5000 -dt 0.1 -vw -cs 3 -cf connectivity_zerlaut_68.zip 
    -ff pearson_0.4_72_-64_-64_19 -gs
\end{verbatim}

A number of simulator settings are exposed to the user on command line. 
Many more settings are exposed to the user but the most important are highlighted. 
The s0..s4 are the flags that set the resolution for the sweep parameters of the specified parameters. 
I.e., '$-s0\ 6$' generates six values equally spaced within the defined range, for the first parameter of the to be explored space. 
Normally the lower and upper bound of the parameters to sweep, can be indicated in the XML-file or can be changed in the driver file. 
This will result in the generation of an array of all the possible parameter combinations, representing the total work-items.
The work-items correspond to the grid-size indicating the total threads spawn on the GPU. 
Each thread spawned is a TVB simulation with a unique parameter combination. 
The two-dimensional grid for the GPU threads is determined recursively according to the number of work-items. 
The program will try to fit the work-items in blocks of 32x32 threads and increase the number of blocks, alternating for the x- and y-axis dimensions in order to find the optimal population of the GPU.

The simulation of the GPU consists of two loops, an inner-loop which determines the number of iterations of the simulation on the GPU and an outer-loop which determines the number of times a GPU-instance is spawn. 
The $-n$ sets the number of simulation steps for the outer-loop and determines the number of times a GPU-instance is spawned.
The steps for the inner-loop are computed by dividing the period of the time-series by the delta-time $-dt$
This influences the precision of the computation of the time-derivatives of the dynamical system, as the end result is averaged of the number of these steps, and reduces the number of times the GPU has to off-load its memory content to disk. 
Next to the spawning of multiple TVB simulations on the GPU itself, the driver makes use of 32 streams for the iterations of the outer-loop, executing the GPU-instances concurrently as well.

\subsection{Output}

After each iteration of the outer-loop, the set of observables is written to memory. 
The number of observables can be manually increased by uncommenting the lines added for the default state-variables or by adding your own observable by setting $[...]$ to a valid expression in the driver file:

\begin{verbatim}
//  tavg(i_node + 0 * n_node) += C_ee/n_step;
    tavg(i_node + 1 * n_node) += [...]/n_step;
\end{verbatim}

The observable is the average of all the inner steps results, for one iteration of the outer-loop.
The connectivity file can be specified with the $-cf$ flag. 
If not specified, a connectome from the standard TVB-library can be selected instead. 
In the example a distinctive connectome, $connectivity\_zerlaut\_68.zip$, originally employed in the scale-integration studies\cite{Goldman2023} is utilized.
The values for the weights of the connectome, used to determine the temporal connections of the regions, are averaged before simulation. 
The $-cs 3$ flag specifies the conductance speed of the connectome in $m/s$ and also determines the depth of the temporal buffer.
The GPU obtains the functional connectivity (FC) by computing the covariance of the resulting time-series for all concurrent TVB simulations, which, in this study, is used for structure to function analysis (FCSC)
The FCSC, the correlation with the structure connectivity (SC), represented by the properties of the connectome and the obtained covariance matrix, is determined with the Pearson correlation. 
The driver can also perform a comparison to an externally obtained reference FC matrix by correlating it to the computed FC of the simulation, also by making use of the Pearson coefficient. 
The file for input can be specified with the $-ff$ flag. 
The framework integrates also the fMRI\_Obs library from \footnote{\url{https://github.com/dagush/fMRI_Observables}}, enabling run-time computation of the Phase Functional Connectivity Dynamics (phFCD) for comparison.
Optionally a Balloon-Windkessel GPU kernel\cite{STEPHAN2007387} can be applied to obtain the BOLD time-series, enabling the comparison with fMRI imaging as well.

\subsection{MPI}

The framework is enhanced with MPI enabling the possibility to deploy multiple parallel processes across compute nodes in a high performance computing cluster.
A single GPU can be assigned to each MPI process and, as there are no dependencies between the execution of the simulations with different parameters, all processes can utilize these resources in parallel.
For example, the command \textit{srun -N2 --ntasks-per-node=4} instantiates multiple MPI processes across 2 nodes, each executing many TVB simulations concurrently on independent GPUs. 
In this example the MPI world consists of $8$ processes, each assigned a single GPU. 
The computational workload, referred to as work-items, will be evenly distributed among all available GPUs. 
This distribution is achieved by reshaping the array in accordance with the world-size and subsequently allocating each rank a corresponding portion of the workload.
When multiple GPUs are specified for execution the driver will sort and gather the results from each instance in the MPI world and output the best ten fittest simulations.
The sorting is enabled using the $-gs$ flag, which is short for grid search.
The $-v$ flag sets the output to verbose and $-w$ saves the best $10$ time-series and their parameters and fitness of each world, to a file. 
When run in verbose mode the driver outputs information about the simulation and GPU memory allocation for rank 0 only. 
It also will also output detailed error information on memory allocation or run-time errors, i.e. too large parameter sets or grid allocation failures. 

The simulation uses a standard forward-Euler method to approximate the time-derivative integral and the linear coupling is applied to globally connect the brain regions for computing the brain dynamics in time, as implemented by the standard TVB-library. 

\subsection{Analysis metrics}
\label{sec:analmeth}

The framework adapts and incorporates an extensive array of tools \cite{Goldman2023,Aquilu2023} as listed in \cref{tab:anali} below in order to analyse the result from the GPU simulations.
The outcomes derived from the analysis are systematically stored in a database using the Python library 'sqlite3,'\cite{sqlite2020hipp} where they are formatted and organized using Structured Query Language (SQL).

\begin{table}[h]
\begin{tabular}{|l|l|}
\hline
\textbf{Metric}          & \textbf{Description}                                                                                                       \\ \hline
mean\_FC                 & Average FC from Pearson corr. of time-series firing rate                                    \\ \hline
mean\_PLI                & Average PLI between brain regions.
\\ \hline
mean\_UD\_duration       & Mean duration of UP and DOWN states                                                                                     \\ \hline
psd\_fmax\_ampmax        & PSD frequency peaks and amplitude                                                              \\ \hline
fit\_psd\_slope          & Fits log(PSD) to log($b/f^a$)                     \\ \hline
\end{tabular}
\caption{The incorporated analysis metrics showing the function name and description.}
\label{tab:anali}
\end{table}

The Phase Locking Index (PLI) is a statistical measure frequently employed in the fields of neuroscience and signal processing. 
Its primary purpose is to evaluate the level of phase synchronization or phase consistency observed between the different nodes of the time series, notably in the context of electroencephalography (EEG) and magnetoencephalography (MEG) data analysis\cite{Goldman2023, silva_pereira_effect_2017, Sazonov2009}.
The PLI serves as a quantification tool for assessing the degree to which the phase angles of two signals exhibit synchronization tendencies over a period of time.

The terms "up" and "down" states are utilized to denote distinct patterns of neuronal activity within the brain. 
These patterns are of particular significance in the realms of neural oscillations and sleep research. 
They are closely linked to the sleep-wake cycle and serve as vital components for comprehending the dynamics of brain function throughout periods of sleep and wakefulness. 

The acronym "PSD" refers to "Power Spectral Density" in the field of signal processing. 
This fundamental concept serves as a means to articulate the manner in which power or energy is distributed within a signal concerning its frequency components. 
PSD plays a pivotal role by furnishing essential insights into the spectral characteristics and frequency content of the signal under examination.

The PSD is approximated by the function: $$log(PSD) = log(b/f^a)$$ 
In which $f$ is the frequency, $a$ and $b$ are the parameters that affects the shape and the amplitude of the function. 
The goal of fitting the PSD to this particular function is to find the values of the parameters $a$ and $b$ that best describe the observed data.  
By applying a logarithm to both sides of the equation a linear relationship is obtained, which is easier to work with mathematically. 
By fitting the PSD data to this function and determining the values of $a$ and $b$, you can gain a better understanding of how the power in the signal is distributed across different frequencies and the characteristics of the spectral density.

\section{Results}

 \subsection{Validation using Functional Connectivity}
To reflect multi-scale brain network activity, biophysical models of the brain can become complex, with dynamical behavior dependent on each parameter as well as potentially nonlinear interactions between parameters. 
Toward progress in understanding the integration of neural phenomena across scales, we have constructed a multi GPU implementation of the TVB-AdEx model; a multi-scale model summarizing the statistics of excitatory and inhibitory populations of spiking neurons \cite{zerlaut2018modeling, divolo} that has demonstrated utility in modeling transitions between conscious- and unconscious-like brain dynamics \cite{Goldman2023, Aquilu2023, goldman2023asynchronous}. 
Specifically, coupling mean-fields representing brain regions by human connectome data \cite{Goldman2023}, it was observed that macroscopic dynamics resembling brain activity emerge during simulation, with global conscious- and unconscious-like activity dependent on parameters representing microscale phenomena known to be related to sleep-wake cycles and the actions of anesthetic agents \cite{Goldman2023, goldman2023asynchronous, Aquilu2023}. 


To further advance beyond previously explored corners of the phase space for the multi-scale TVB-AdEx model with the published implementation \cite{Aquilu2023, Goldman2023, goldman2023asynchronous}, prohibitively large computational resources would have been required. 
Therefore, we have implemented the TVB-AdEx model for GPU, which can be more efficient for computations that are embarrassingly parallel. Parameterization in general is embarrassingly parallel due to an absence of dependency between simulations; each parametric combination does not depend on other concurrent simulations. GPU can therefore be an excellent choice for acceleration; the data latency of the individual simulations is largely covered by the many concurrent simulations. 

For detailed description of the implementation of our TVB-HPC framework, see materials and methods. 
Briefly, RateML tools were modified and manually annotated to produce the simulation code, its dependencies and outputs, including the transfer function and analysis pipelines. 
The framework can be initiated on any CUDA capable GPU device.

To determine whether the GPU model robustly reproduces the behavior of previously published CPU implementation \cite{Goldman2023, Aquilu2023, goldman2023asynchronous} the FC of four simulations with different parameter-sets, corresponding to different states of the model, are compared to the online computed FC from the simulation of the TVB-AdEx model on the GPU, in order to find the best matching sets of parameters reproducing the model's behaviour. 
Parameter-sets with values $S_0(g=0.0, b_e=0.0), S_1(g=0.3, b_e=24), S_2(g=0.4, b_e=72), S_3(g=0.4, b_e=120)$ are considered.
For full description on how to make use of this comparison, see the material and methods section.
The fitness is defined as the pearson correlation between the external and the computed FC computed with \cref{eq:pearsson}. 

\begin{equation}
 \rho_{xy} = \frac{\mathrm{Cov}(x,y)}{\sigma_{x}\sigma_{y}} \, , 
 \label{eq:pearsson}
\end{equation}

\begin{figure}[htbp]
  \begin{minipage}{0.4\textwidth}
    \vspace{1.5cm}
    \begin{table}[H]
        \hspace{.7cm}
      \begin{tabular}{|c|c|c|c|c|}
        \hline
            \textbf{g} & $\mathbf{b_{\textbf{e}}}$  & $\mathbf{E_{\textbf{L,e}}}$  & \textbf{Fit} & \textbf{Pos} \\ \hline
            0          & 0             & -64          & 0.62         & 1541         \\
            0.8        & 0             & -60          & 0.73         & 1            \\ \hline
            0.3        & 24            & -64          & 0.81         & 12           \\
            0.9        & 120           & -64          & 0.86         & 1            \\ \hline
            0.4        & 72          & -70          & 0.77         &  58            \\
            0.84       & 113           & -64          & 0.79         & 1            \\ \hline
            0.4        & 120           & -64          & 0.78         & 15           \\
            0.8        & 120           & -64          & 0.82         & 1            \\ \hline
      \end{tabular}
      \vspace{1.cm}
      \caption{\textbf{
      Parameters combinations with the largest fitness, comparing FC derived from CPU for four different parameter explorations with GPU-implemented simulations.} 
      Each pair represents the fitness of the combination searched for and the highest fitness out of $15,625$ combinations. 
      Overall a fitness of >70\% is achieved, indicating that the GPU model displays similar behaviour to the original model from the TVB library used in the CPU implementation to reproduce results from previous studies of the TVB-AdEx.}
      \label{tab:sweep_results}
    \end{table}
  \end{minipage}%
  \hfill
  \begin{minipage}{0.44\textwidth}
    \hspace{-1.6cm}
    \includegraphics[width=9.0cm]{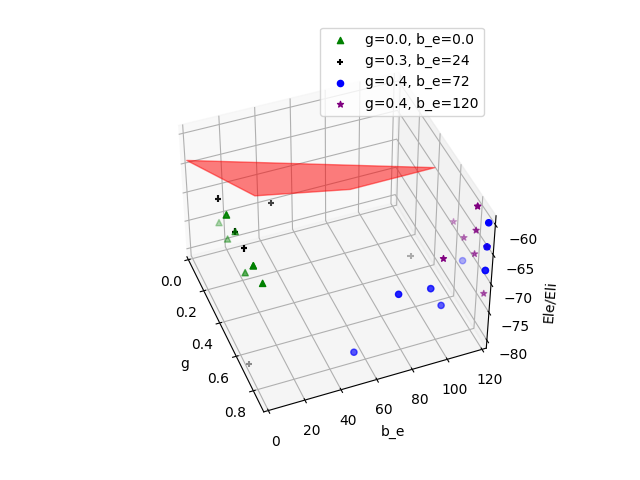}
    \caption{\textbf{Visual representation of the best 10 solutions of the external FC in parameter space.} 
    The corners of the plane depicted in red, are the four coordinates representing the solutions for the fitting.
    Each symbol has 10 solutions plotted.
    The distance to the plane of the data points represents the fitness; the larger the fitness, as depicted in \cref{tab:sweep_results}, the closer the solutions are to one of the corners of the red plane. 
    }
    \label{fig:diffFC}
  \end{minipage}
\end{figure}

The results of the four FC comparisons, conducted on data from a simulation of the model for different states, are presented in \cref{tab:sweep_results}.
In total $15,625$ parameter-combinations are considered; each parameter has a resolution of 6. 
Each tuple in the table provides two rows of information: the first row displays the values for the parameters to be found and their fitness value and position, while the second row shows the highest fitness achieved and the corresponding matching parameters. 
I.e. the first row indicates that the FC with parameters $g=0.4, b_e=80, E_{l,e}=~E_{l,i}=-70$ is found at position $58$ out of $15,625$ with an accuracy of 77\%. 
The best fitness for that comparison yielded the parameters value $g=0.84, b_e=113, E_{l,e}=~E_{l,i}=-64$ with an match accuracy between parameters for CPU and GPU implementations of 79\%.
As is displayed, the values for the parameters are often found with high accuracy, but overall the GPU model prefers higher coupling parameter values. 

The results in \cref{fig:diffFC} provides a visual representation of the distribution of the top 10 distinct solutions across the parameter space for the different parameter-sets.
In order to better visualize the solution space, a plane has been created with the four individual solutions as its corners. 
The smaller the distance to one of the corners of the red plane, the better the match with the particular parameter-set between CPU and GPU implementations.
The results indicate an overall parameter match achieving approximately 70\% similarity with the reference FC. 
Notably, the $E_{le}$ and $E_{li}$ parameters exhibit slight differences, aligning with a prediction made by the model\cite{Aquilu2023}. 
Therefore, only the first parameter is displayed in the table and figure.

\subsection{Vast parameter space exploration and analysis}
Fully leveraging the computational resources of the Juwels Booster compute cluster at the Forschungszentrum Jülich empowers extensive parameter space exploration, enabling a comprehensive investigation across a wider spectrum of parameters, as described in \cref{tab:vastpse} in \cref{sec:analmeth}. 
The extensive analysis capability has allowed us not only to validate the model, but also to investigate the extensive properties of the parameter space associated with this model. 
It facilitates the search for transitions in simulated brain dynamics, identification of peaks in the power spectrum, characterization of Up and Down states typical for synchronous behavior (for full description, see \cref{tab:anali}), and examination of factors such as the influence of the velocity of action potential propagation through the connectome on adaptation (for full description, see \cref{tab:vastpse}).

\begin{table}[]
\begin{tabular}{|l|l|l|l|}
\hline
\textbf{Name}     & \textbf{Range}                  & \textbf{Resolution} & \textbf{Description}                                               \\ \hline
$g$     & {[}.1, .9{]}           & 8          & Coupling strength connectome                               \\ \hline
$b_e$         & {[}0., 100.{]}         & 8          & Spike-frequency adaptation [pA]                                   \\ \hline
$weight Noise$ & {[}1e-9, 1e-4{]}       & 4          & Scaling weight of noise                        \\ \hline
$speed$        & {[}1., 7.{]}           & 4          & Connectome Speed [m/s]                              \\ \hline
$\tau_{w} $   & {[}250, 750{]}         & 4          & Adaptation time constant exc. neurons [ms] \\ \hline
$a_e $        & {[}-10.0, 20.0{]}      & 4          & Sub-threshold adaptation conductance [nS]             \\ \hline
$c_{F_{e,e}}$      & ${[}.3e^{-3}, .5e^{-3}{]}$ & 2          & External input [Hz]                                \\ \hline
$c_{F_{e,i}} $      & ${[}0, .5e^{-3}{]}$ & 2          & External input [Hz]                                \\ \hline
$c_{F_{i,e}}$       & ${[}.3e^{-3}, .5e^{-3}{]}$ & 2          & External input [Hz]                            \\ \hline
$c_{F_{i,i}} $      & ${[}0, .5e^{-3}{]}$ & 2          & External input [Hz]                            \\ \hline
\end{tabular}
\caption{Parameters targeted for vast exploration.}
\label{tab:vastpse}
\end{table}

\subsubsection{Effect of modulating coupling and adaptation strength}
\begin{figure}[H]
\includegraphics[width=15.7cm]{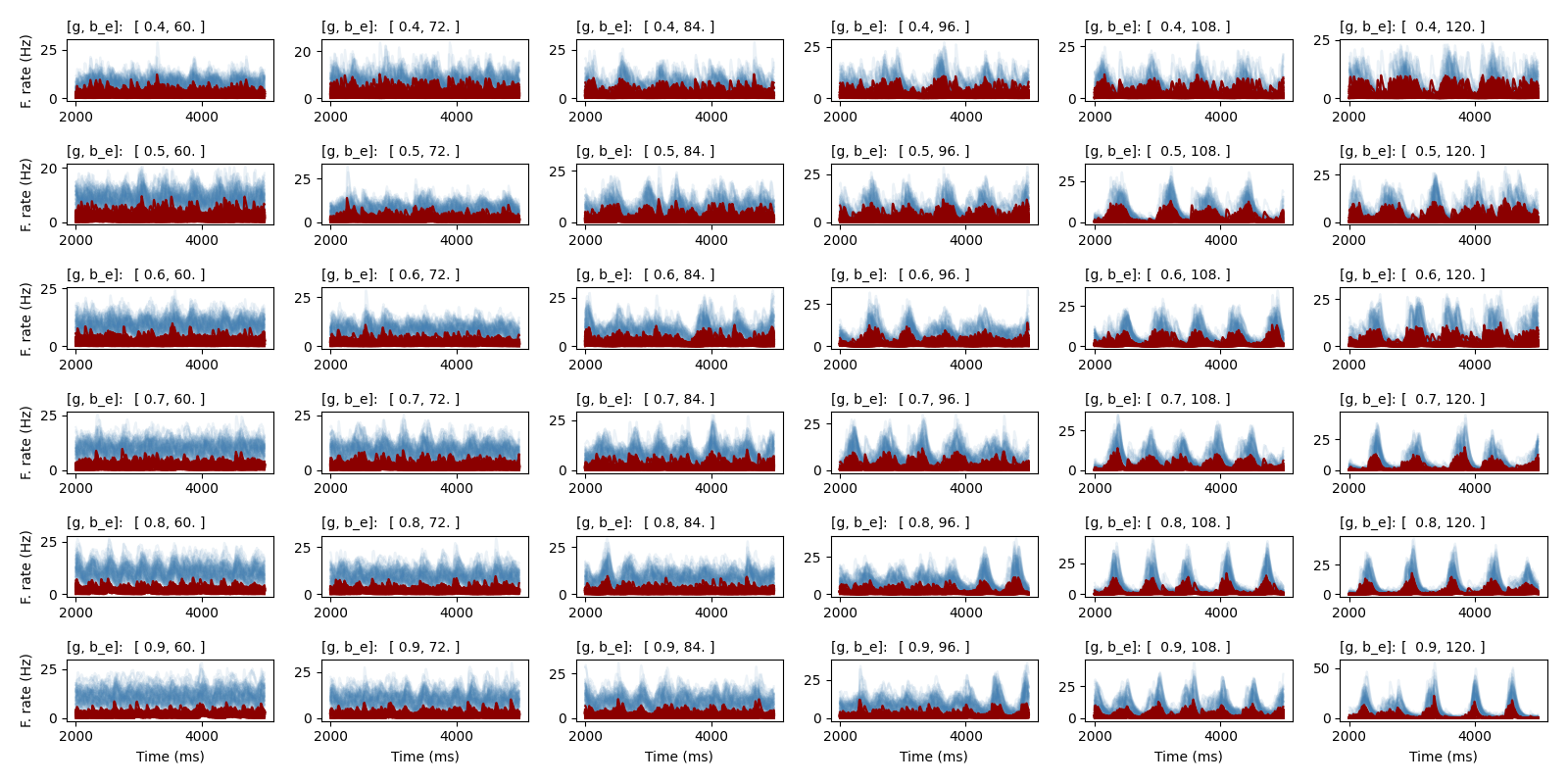}
\caption{\textbf{Results of a 6x6 parameters sweep for the global coupling, $g$, and spike-frequency adaptation, $b_e$, parameters.} 
The exhibitoy firing rate, $F_e$, is depicted in red and the inhibitory firing rate, $F_e$, in blue.
The $g$ and $b_e$ parameters are swept for ranges $[.4, .9]$ and $[60, 120]$ respectively. 
Results show the output time-series for each individual set of parameters printed above each plot. 
The plots show the interaction of $g$ and $b_e$, resulting in a gradual transition from asynchronous (top left corner) to synchronous behavior (bottom right corner) for the GPU TVB-AdEx model.
\label{fig:allsweep}}
\end{figure}

The TVB-AdEx model can be used to study different types of brain dynamics associated with different states of consciousness; the asynchronous irregular dynamics associated with wakeful, conscious awareness and synchronous, regular dynamics associated with unconscious brains states. 
To validate whether the GPU model can represent these regimes, the global coupling, $g$, and spike-frequency adaptation, $b_e$, variables previously studied \cite{Aquilu2023} were swept over the aforementioned ranges (table 4). 
In \cref{fig:allsweep}, time-traces display the raw output for a sweep of $g$ and $b\_e$ in the ranges of $[.4, .9]$ and $[60, 120]$ respectively for a total of $36$ parameter combinations, showing the model's ability to simulate transitions between disordered asynchronous to more ordered synchronous behavior. 
These results demonstrate the interaction between coupling and spike-frequency adaptation, showing most synchronous-regular activity in the bottom right corner panel, where both parameters are maximal, thus further replicating previously established behavior of the CPU model using GPUs for the TVB-HPC framework. 

\begin{figure}[H]
\includegraphics[width=11.4cm]{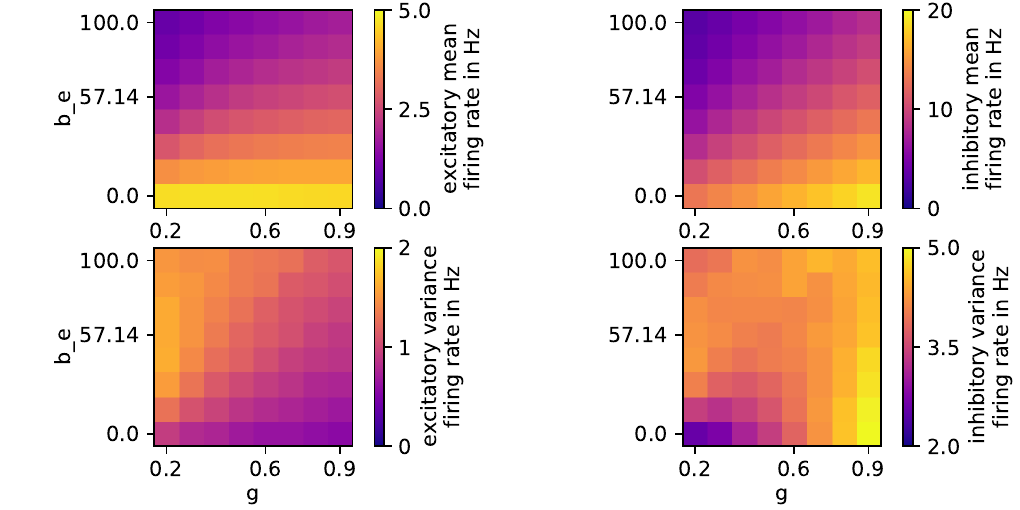}
\caption{Parameter exploration for global coupling and spike-frequency adaptation.
The other parameters are: $weight\_Noise: 1.0e^{-06}, speed: 3.0, \tau_{w}: 417, a_e: 0.0, c_{F_{e,e}}: 0.3e^{-3}, c_{F_{e,i}}: 0.0, c_{F_{i,e}}: 0.3e^{-3}, c_{F_{i,i}}: 0.0$
\label{fig:exploration_g_be}}
\end{figure}

Furthermore, \cref{fig:exploration_g_be} shows that increasing coupling reduces the variability of the firing rate of $F_e$ while increasing the mean $F_i$. In contrast, increasing adaptation strength increases the variance of $F_e$.
These results concur with the plotted time-series in \cref{fig:allsweep}, showing that for high values of coupling with low values of adaptation strength (bottom left) the variance of $F_e$ is low, but $F_i$ is high. 
Visa versa, the variance is high for both when the adaptation strength is high, but the coupling is lower. 
Furthermore, a notable presence of a bifurcation is observed along the diagonal of the parameter space defined by $g$ and $b_e$ for $F_i$.

\subsubsection{Effect of modulating the adaptation time constant}

\begin{figure}[H]
\includegraphics[width=\textwidth]{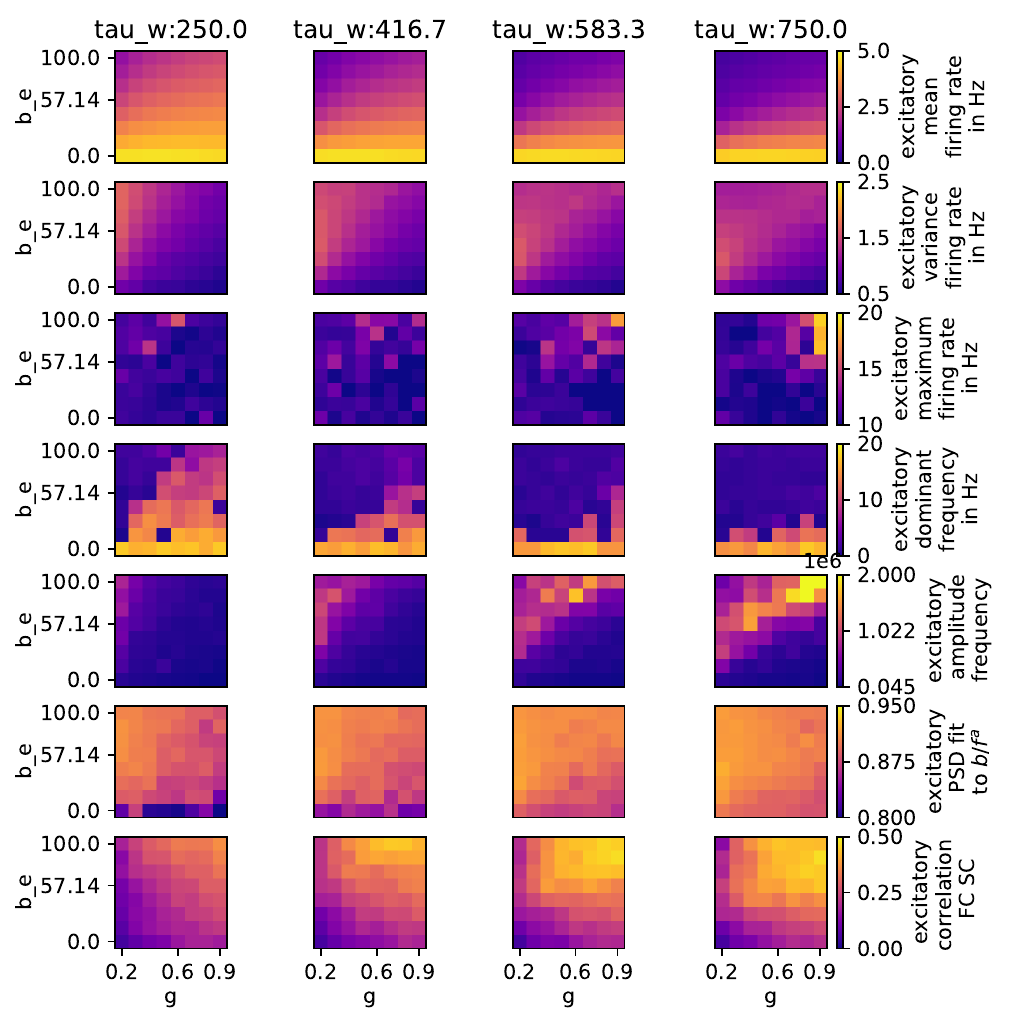}
\caption{Parameter exploration for global coupling and adaptation strength when modulating the adaptation time constant $\tau_{w}$. 
The other parameters are: $weight\_Noise: 1.0e^{-06}, speed: 3.0, a_e: 0.0, c_{F_{e,e}}: 0.3e^{-3}, c_{F_{e,i}}: 0.0, c_{F_{i,e}}: 0.3e^{-3}, c_{F_{i,i}}: 0.0$ 
\label{fig:exploration_tau_w}}
\end{figure}

As depicted in \cref{fig:exploration_tau_w}, an increase in the adaptation time parameter $\tau_{w}$ yields several effects within the model. 
Specifically, it leads to a reduction in the mean firing rate, enhances the linearity of the power-frequency relationship, diminishes the peak frequency value, and promotes an increase in correlation with the structural connectivity (SC).

\subsubsection{The effect of modulating excitatory subthreshold adaption conductance}

The model exhibits activity only under conditions of low noise when two specific criteria are met: firstly, when the action potential speed is equal to $1.0$ m/s (as indicated before), and secondly, when the value of $a_e$ is not equal to $-10$ pA. 
The occurrence of activity at $a_e$ = $-10$ pA is attributed to the fact that, during this particular condition, the adaptation is acting as an excitatory mechanism, due to negative values.


When $a_e$ reaches $-10$, an issue arises where excitation becomes excessively pronounced, resulting in all regions exhibiting excitatory activities exceeding $100$ Hz, a level deemed excessively high and potentially related to paroxysmal activity and seizure dynamics. 
However, the model lacks precision when dealing with high firing rate activity and lacks an explicit mechanism for transitioning between high and low activity states.
The parameter $a_e$ introduces a bifurcation effect, specifically impacting the mean measurements as it fluctuates between higher and lower values at $a_e = -10$. 
This effect is distinct from the bifurcation observed with parameters $b_e$ and $g$, which occur at higher values of $b_e$.

Furthermore, increasing the value of $a_e$ leads to several outcomes, including a reduction in firing rates, decreased variability, diminished impact of parameter $b_e$, a decrease in the dominance of certain frequency components, and an increase in the correlation between different regions.

\begin{figure}[H]
\includegraphics[width=14cm]{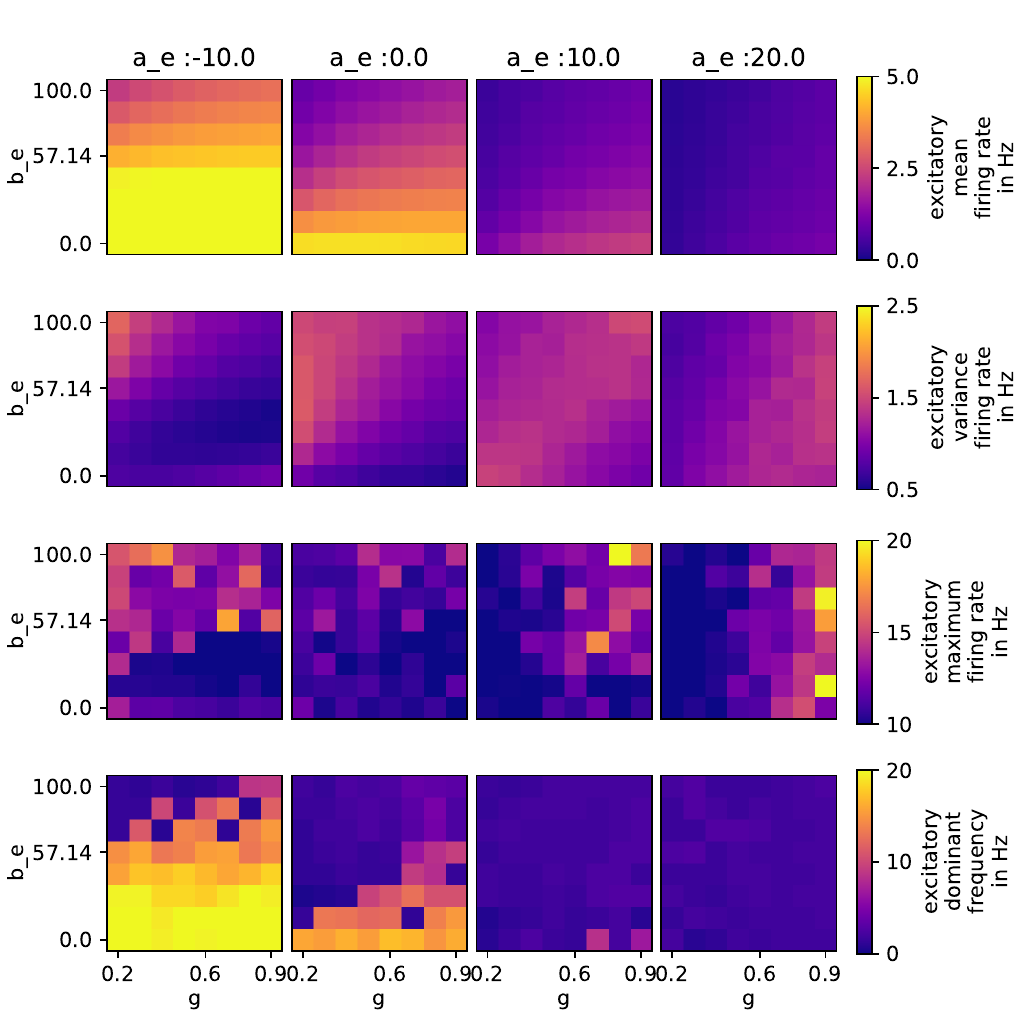}
\caption{Parameter exploration for global coupling and adaptation strength when modulating the subthreshold adaptation conductance, $a_e$. 
The other parameters are: $weight\_Noise: 1.0e^{-06}, speed: 3.0, \tau_{w}: 417, c_{F_{e,e}}: 0.3e^{-3}, c_{F_{e,i}}: 0.0, c_{F_{i,e}}: 0.3e^{-3}, c_{F_{i,i}}: 0.0$ 
\label{fig:exploration_a_e}}
\end{figure} 

\subsubsection{The effects of stimulating external excitatory and inhibitory populations}

\begin{figure}[H]
\includegraphics[width=14cm]{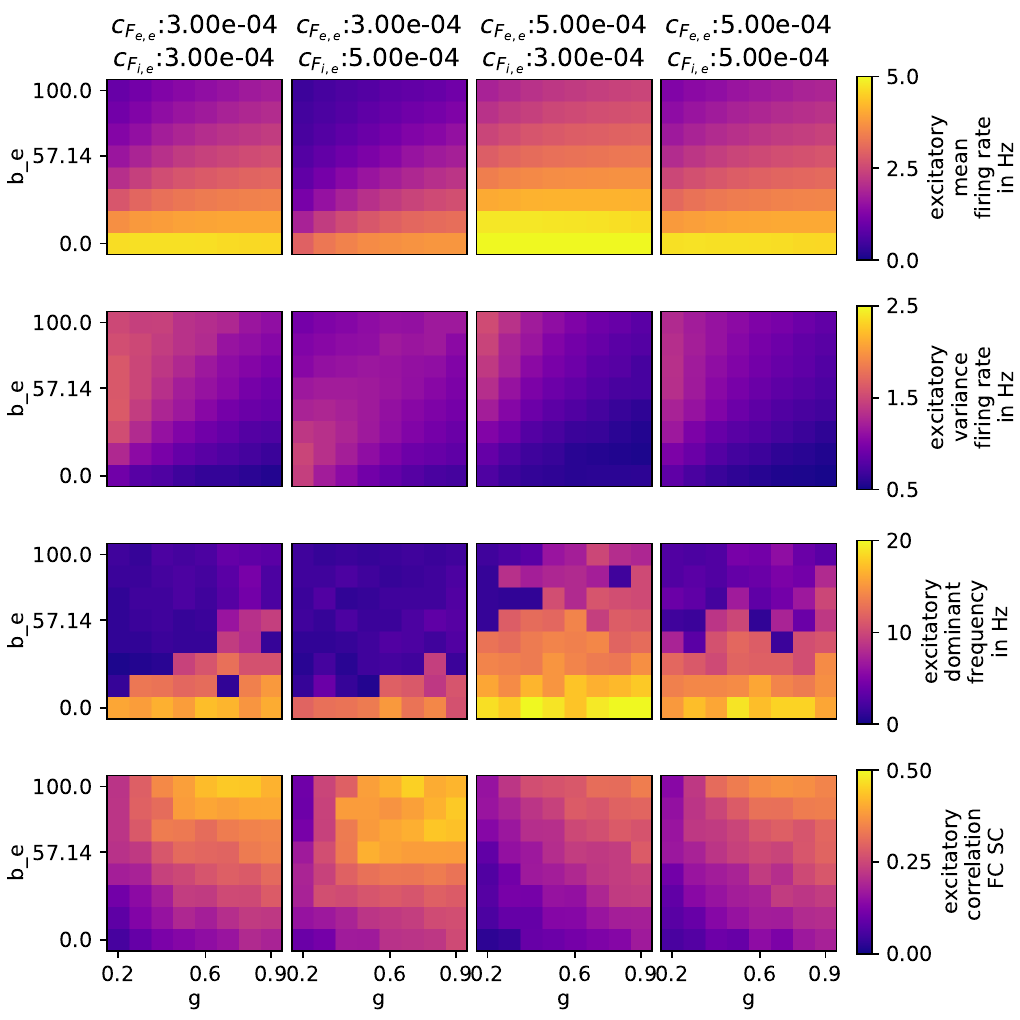}
\caption{Parameter exploration for global coupling and spike-frequency adaptation for different external excitatory input.
The other parameters are: $weight\_Noise: 1.0e-06, speed: 3.0, \tau_{w}: 417, a_e: 0.0$
\label{fig:exploration_exin}}
\end{figure}

As depicted in \cref{fig:exploration_exin}, in the context of the model's dynamics, stimulating the excitatory population leads to a shift in mean activity levels, causing transitions both upward and downward, while concurrently reducing the correlation with the structural connectivity (SC). 
Conversely, stimulating the inhibitory population results in an increase in the mean excitatory firing rate and a decrease in the global spectral frequency. 
It is important to note that the excitation of both the excitatory and inhibitory neuronal populations collectively influence the maximum peak of variance within the excitatory population.

In the context of synchronization analysis, two distinct conditions with specific criteria are identified:
\\\\
\textbf{Condition 1}: High Mean Phase Lag Index (PLI) in the excitatory population
For this condition, parameters are considered if they meet the following criteria:

\begin{itemize}
    \item Excitation of the excitatory population ($c_{F_{e,e}} = 0.5e-3$).
    \item Low $a_e$ value ($-10$).
    \item Low $\tau_{w}$ value ($250$).
    \item Speed higher than $1.0$ m/s.
    \item Low weight noise ($< 1.0e-5$).
\end{itemize}

With these criteria, $1,152$ parameters are considered, resulting in $927$ cases where $PLI_e > 0.33$ and $991$ cases where $PLI_i > 0.3$. 
\\\\
\textbf{Condition 2}: High Mean Functional Connectivity (FC) in the excitatory population. 
This condition includes parameters that meet the following criteria:

\begin{itemize}
    \item Low weight of noise ($1.0e-7$).
    \item Low $a_e$ value ($-10$).
    \item Speed higher than $1.0$ m/s.
    \item Requirements for $\tau_{w}$ and excitation of the excitatory population:
        \begin{itemize}
            \item Either low $\tau_{w}$ and low excitation of the excitatory population ($\tau_{w} = 250$ and $c_{F_{e,e}} = 0.3e-3$).
            \item Or high $\tau_{w}$ and high excitation of the excitatory population ($\tau_{w} != 250$ and $c_{F_{e,e}} = 0.5e-3$).
        \end{itemize}
\end{itemize}

With these criteria, $1,536$ cases are considered, with 1152 satisfying $FC_e > 0.98$ and additionally, $5$ cases outside these conditions were found. 

\subsubsection{Pathology and the effect of modulating the propagation speed of action potentials through the connectome}

One intriguing aspect of mean field models obtained through biophysical approaches is their potential utility in identifying abnormal patterns or irregularities associated with pathologies.
The TVB-AdEx can be used to investigate seizure activity related to Epilepsy for instance, due to hyperactive or hypersynchronized states of the model\cite{Goldman2023}. 
The dynamical landscape of the model comprises a pathological fixed point around 190Hz\cite{Aquilu2023}. 
This is a point where neurons fire again just after their refractory period. 
The vast parameter sweep can be used to chart the landscape and find the conditions under which the mean-fields of the network settle at this pathological fixed point. 
The occurrence of reaching this fixed point can be localized by examining the conditions for a jump in the activity to high frequency activity.  
Disconnected models and models with a lower global coupling have a higher probability reaching this fixed point. 
It is to be expected that the fixed point is less likely to be reached if the coupling is stronger.

\begin{figure}[H]
\includegraphics[width=\textwidth]{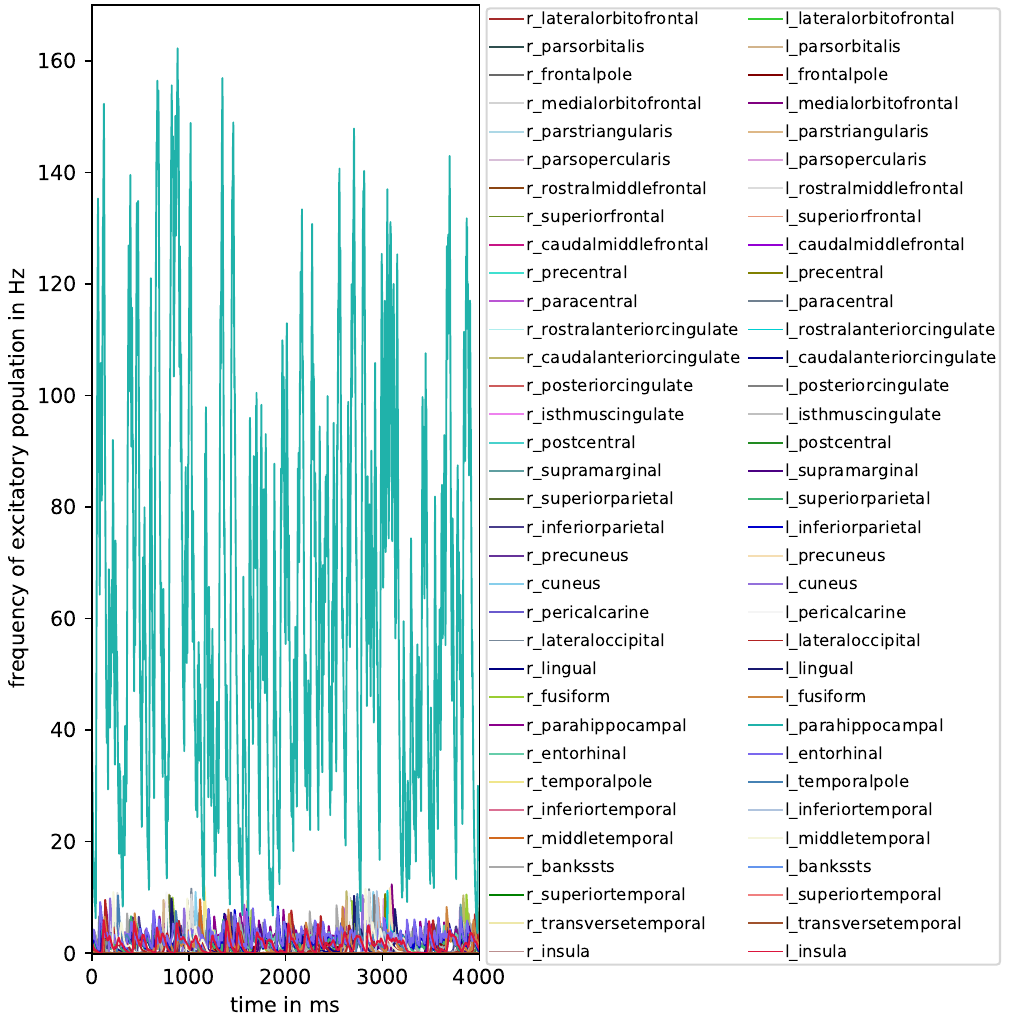}
\caption{Time-series of the excitatory population firing rate for each brain region when action potential velocity is equal to 1 m/s and $g: 0.4, b_e: 86, weight\_Noise: 1e^{-06}, \tau_{w}: 417, a_e: 0.0, c_{F_{e,e}}: 0.3e^{-3}, c_{F_{e,i}}: 0.0, c_{F_{i,e}}: 0.3e^{-3}, c_{F_{i,i}}: 0.0$
}
\label{fig:timeserrie_spee_1}
\end{figure}

Examining the effects of modulating the speed of action potential propagation through the connectome, it becomes apparent that when the speed attains a value of $1.0$ m/s, particular regions display varying probability to visit the paroxysmal fixed point and display high firing rate activity exceeding 100 Hz. 
This observation is visually depicted in \cref{fig:timeserrie_spee_1}.
When quantifying the instances in which a high firing rate is observed exclusively in single regions, we observe the following.
The left parahippocampal region visits the paroxysmal fixed point, for all $16,384$ parameter combinations when the speed attains a value of $1.0$ m/s. 
Furthermore, the following counts are recorded: 325 occurrences for the left enthorhinal region, 17 occurrences for the left infereriortemporal region, 1 occurrence for the right parsopercularis region, and 1 occurrence for the right isthmuscingulate region.
Intriguingly, the left parahippocampal region stands out with the highest input weights and has the strongest connection with the left enthorinal region. 
This observation may suggest that highly connected regions could be implicated in the generation of epileptic seizures.
This specific observation is only found when the speed property of the connectome is low and affirms the assumption that less connected brains have a higher probability of reaching the pathological fixed point. 

\begin{figure}[H]
\includegraphics[width=\textwidth]{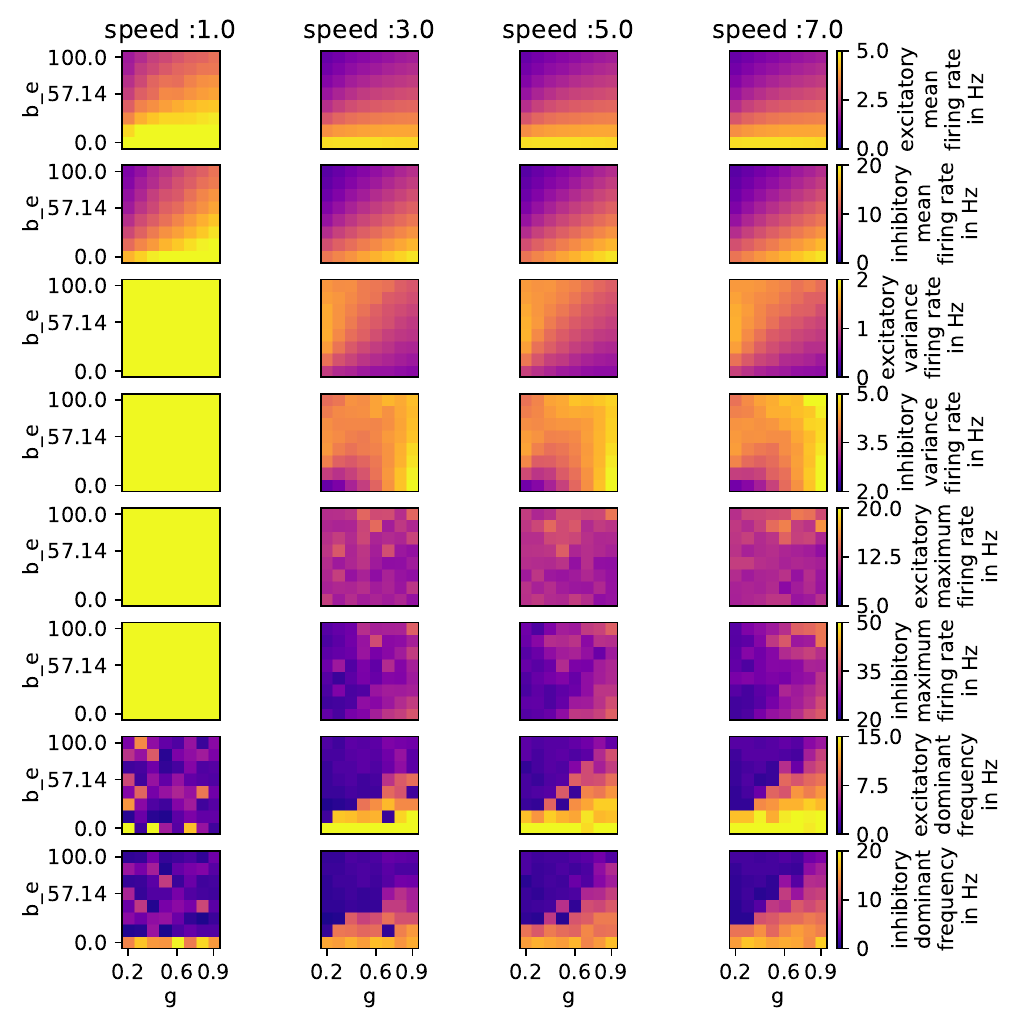}
\caption{Parameter exploration for coupling and adaptation strength of excitatory neurons when action potential propagation speed is modulated. 
The graphs depicted in the speed: 1.0 column, indicate  divergent behavior in this region of parameter space. 
The other parameters have the values: $weight\_Noise: 1.0e^{-06}, \tau_{w}: 417, a_e: 0.0, c_{F_{e,e}}: 0.3e^{-3}, c_{F_{e,i}}: 0.0, c_{F_{i,e}}: 0.3e^{-3}, c_{F_{i,i}}: 0.0$
}
\label{fig:exploration_speed}
\end{figure}

Apparently, the impact of speed is not significant unless it falls below or exceeds the value of $1.0$ m/s.
In \cref{fig:exploration_speed} this is once more affirmed, in which the coupling and adaptation strength are modulated each with a different value for speed. 
These results indicate that only for a speed of 1.0 m/s the maximum and variance in firing rates for both excitatory and inhibitory populations is maximal. 
When the speed increases, the firing rate does not exceed 20 Hz, which is similar to healthy brain dynamics. 
The first two graphs, the mean firing rate for both population, of the speed: 1.0 column, also suggests that not all regions demonstrate an inclination to reach such elevated firing rates.

\subsection{Performance}

In a previous study the scaling relation between a CPU and GPU implementation has already been performed\cite{rateml2022}.
The reported CPU implementation also makes use of the TVB-numba backend but implements the Montbrio-Paxin-Rosin model\cite{Montbrio2015}, of which the scaling results are comparable to the TVB-AdEx CPU implementation. 
In this section we report the execution time for the GPU simulation and analysis for the AdEx model only. 

\begin{figure}[H]
\includegraphics[width=14 cm]{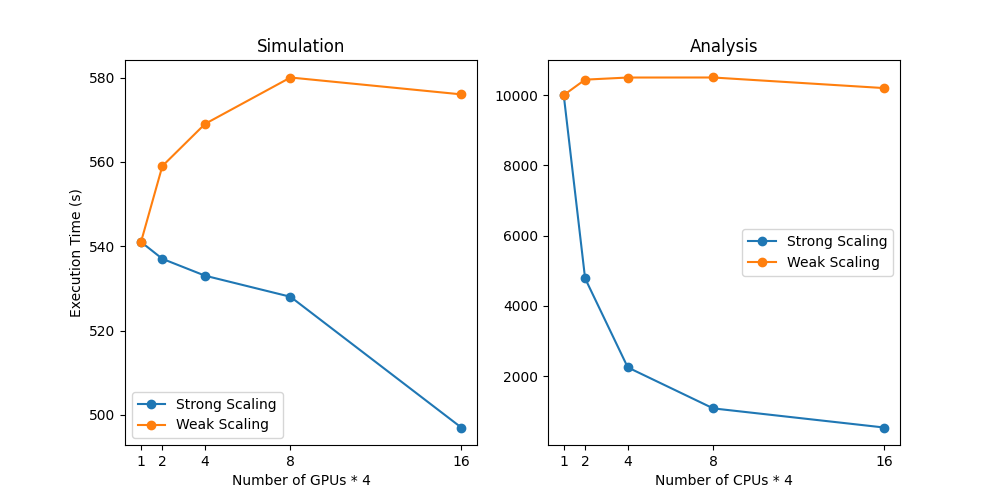}
\caption{\textbf{Scaling resources} 
Strong and weak scaling results for the framework with AdEx model for the simulation and analysis part of the framework doubling the number of resources. 
The x-axis displays the number of nodes each equipped with four GPUs and CPUs.
The results are obtained on the Juwels Booster cluster located at the Forschungszentrum Juelich.
\label{fig:sw_scaling}}
\end{figure}

The execution time analysis, plotted in\cref{fig:sw_scaling}, indicates linear results for strong and weak scaling in relation to doubling the number of GPUs. 
The first graph displays the scaling behaviour for the simulation part, for which the overall execution time remains under $600$ sec. 
There is a slight decrease in execution time when strong scaling the simulation part of the framework. 
This is due to the fact that the GPU is already optimal when fully occupied, making use of many threads to hide memory latency.
Adding more GPUs and dividing the load, makes the execution relatively less efficient because of a lesser thread count per GPU to hide latency.

The second graph shows the execution time for the analysis part of the framework, which is conducted on the CPUs of the nodes. 
As expected, the analysis part benefits greatly from adding more resources, as run-time diminishes by half every time resources are doubled. 
The weak scaling indicates that the execution time for the analysis, which is the bottleneck of the framework, remains at a constant at approximately $10,000$ seconds. 

The temporal buffer maintains a record of the model states throughout the progression of a TVB simulation, incorporating signal propagation delays. 
The size of the temporal buffer is determined by the product of the number of states, the maximum length of the connectome, and the number of brain regions. 
Consequently, the temporal buffer constitutes the primary contributor to the GPU's memory footprint, rendering the GPU's scalability and maximum number of parameters contingent upon either the number of model states or the size of the connectome.

The maximum number of resources of the cluster utilized for simulations is $384$ nodes (of 936 nodes), which have each $4$ GPUs, totalling $1536$ GPUs.
This amount of resources enabled a parameter exploration of $25,165,824$ concurrent TVB instances, each running simulations with different set of parameters. 
The execution time for handling such a vast number of instances, encompassing both the simulation and analysis phases, remains well below the thresholds of $530$ seconds and $10,000$ seconds, respectively.

\section{Discussion}

In this work we report the GPU implementation of the TVB-AdEx mean-field model used in studies on consciousness to link mechanisms operating at microscopic scales to global brain dynamics. 
We explore the potential enhancements achievable through the utilization of MPI for vast GPU cluster parallelization and show the possibilities of increasing computational resources for complex brain models.
The data generated from numerous parameter combinations can become overwhelming. 
We demonstrate that by conducting the analysis directly and organizing the outcomes in a database format, this framework enables oversight and becomes highly reusable. 
Furthermore, its modular architecture allows users to incorporate additional mean-field models for distributed processing across a compute cluster, especially for resource-intensive analyses. 

Results on previously implemented models in the GPU framework, such as the Kuramoto\cite{kuramoto1975international}, Wong-Wang\cite{wong2006recurrent}, the Epileptor\cite{jirsa2014nature}, and Montbrio-Paxin-Rosin\cite{Montbrio2015}, showing performant behaviour have been previously reported\cite{rateml2022}.
With this work we go one step further by providing a strategy to address more complex models, like the AdEx mean field model, which can largely benefit from optimization on GPUs.
The capability to compare the simulated output with (dynamical) functional connectivity renders this framework highly suitable, for instance, in the creation of a digital twin. 
It endows the user with the capacity to discern connections between externally acquired data, whether empirical or simulated neuroimaging data like EEG, or even an externally obtained BOLD signal through the utilization of the GPU BOLD kernel. 


As documented in\cite{Aquilu2023}, the multi-node implementation employing MPI for the purpose of exploring a constrained parameter space in this model allocated all unique parameter combinations or work-items to a single core out of the 128 cores available on the JUSUF supercomputer in Jülich, taking roughly 6 minutes of run-time for 5 seconds of simulated brain time. 
The GPU-AdEx model takes about $8$ minutes to compute $5$ seconds of biological time, for 16384 work-items concurrently; increasing the resolution of the parameter space by 128 times for the simulation part of the framework. 
Upon scrutinizing the scalability of the implemented analysis tools, it becomes evident that this will potentially benefit from a GPU implementation also. 
This is exemplified by the observation that doubling the computational resources results in a reduction of run-time by half.
Such an implementation would be well-suited for the embarrassingly parallel characteristics of the analysis pipelines, accelerating the reconnaissance of the parameter space even further.  

The Juwels Booster cluster at the Juelich Forschungszentrum has a total of $936$ nodes 
This substantial resource pool allows for the theoretical exploration of parameter spaces that encompasses a staggering $61,341,696$ possible configurations, all within a timeframe comparable to our current execution times.
These impressive computational resources usher in a new era, one in which every single parameter of the model can be comprehensively examined, leaving no aspect unexplored.  
However, it's worth noting that new challenges and limitations emerge in this expanded landscape. 
For instance, the amount of memory required to store the vast amounts of information for plotting and analysis becomes a crucial consideration. 

Additionally, the sheer volume of data generated necessitates more sophisticated methods of data exploration and analysis, surpassing the capacity of a single individual to manage effectively. 
In this regard, the utilization of specialized tools becomes a necessity. 
One such tool that holds promise in navigating these immense datasets is "Learning to Learn" (L2L)\cite{yegenoglu2022}. 
This automated machine learning framework is purpose-built for high-performance computing environments and is adept at employing gradient or evolutionary strategies to traverse the vast data generated by this framework.
Furthermore, the TVB-HPC framework can serve as an initial step in the process of mapping the intricate properties of the parameter space. 
These preliminary insights can then be leveraged by Learning to Learn to intelligently identify and elucidate intriguing patterns within this expansive and complex space.

In this work we utilized an HPC implementation of analysis toolkits, taking a step in the direction of creating a complete workflow specifically designed to address metrics related to transitions in brain dynamics associated with varying degrees of consciousness. 
Thus, this comprehensive simulation and analysis framework distinguishes itself as a unique and invaluable tool for advancing the development of digital twins, not only within the realm of theoretical research but also in the context of personalized clinical investigations for epilepsy, sleep, anesthesia, and disorders of consciousness.  


\vspace{6pt}

\section*{Author Contributions}
Conceptualization, Michiel van der Vlag, Sandra Diaz and Jennifer Goldman; \\
Data curation, Michiel van der Vlag and Sandra Diaz; \\
Formal analysis, Michiel van der Vlag, Lionel Kusch and Jennifer Goldman; \\
Funding acquisition, Alain Destexhe, Viktor Jirsa and Sandra Diaz; \\
Investigation, Michiel van der Vlag, Lionel Kusch and Jennifer Goldman; \\
Methodology, Michiel van der Vlag, Lionel Kusch, Sandra Diaz and Jennifer Goldman; \\
Project administration, Michiel van der Vlag, Alain Destexhe, Viktor Jirsa and Sandra Diaz; \\Resources, Alain Destexhe, Viktor Jirsa and Sandra Diaz; \\
Software, Michiel van der Vlag and Lionel Kusch; \\
Supervision, Alain Destexhe, Viktor Jirsa, Sandra Diaz and Jennifer Goldman; \\
Validation, Michiel van der Vlag and Lionel Kusch; \\
Visualization, Michiel van der Vlag and Lionel Kusch; \\
Writing – original draft, Michiel van der Vlag, Lionel Kusch and Jennifer Goldman; \\
Writing – review \& editing, Michiel van der Vlag and Jennifer Goldman.

\section*{Funding}
The research leading to these results has received funding from the European Union’s Horizon 2020 Framework Programme for Research and Innovation under the Specific Grant Agreements No. 785907 (Human Brain Project SGA2) and 945539 (Human Brain Project SGA3). This research has also been partially funded by the Helmholtz Association through the Helmholtz Portfolio Theme "Supercomputing and Modeling for the Human Brain". 

\section*{Acknowledgements}
The authors thank David Aquilue, Nuria Tort-Colet, and Trang-Anh E. Nghiem for contributing analysis scripts. 
We thank David Aquilue also for helpful discussion of the manuscript in progress and for contributing data from his previous work \cite{Aquilu2023} for comparison with the GPU AdEx model reported here.

\section*{Conflics of interest}
The authors declare no conflict of interest.

\bibliographystyle{plain}
\bibliography{biblio}

\end{document}